\documentclass[11pt]{myarticle}
\usepackage{mycite}
\usepackage{latexsym}			
\renewcommand\refname{}		
\usepackage{graphicx}
\usepackage{color}
\definecolor{black}{rgb}{0,0,0}
\definecolor{red}{rgb}{1,0,0}
\definecolor{darkblue}{rgb}{0,0,0.7}
\newcommand\bla{\color{black}}
\newcommand\red{\color{red}}

\newcommand\go{\mathrel{\raise.3ex\hbox{$>$}\mkern-14mu 
             \lower0.6ex\hbox{$\sim$}}}
\newcommand\lo{\mathrel{\raise.3ex\hbox{$<$}\mkern-14mu 
             \lower0.6ex\hbox{$\sim$}}}
\begin{document}

\sf 

\huge
\noindent
\textbf{%
Ring dynamics around non-axisymmetric bodies
}%
\normalsize

\vspace{0mm}

\noindent
\textbf{%
B. Sicardy$^{*,1}$,
R. Leiva$^{2}$,
S. Renner$^{3}$,
F. Roques$^{1}$,
M. El Moutamid$^{4,5}$,
P. Santos-Sanz$^{6}$,
J. Desmars$^{1}$
\sf [This is a pre-reviwed version of an letter published in Nature Astronomy.
The final authenticated version is available inline at 
http://dx.doi.org/10.1038/s41550-018-0616-8 and \\
https://rdcu.be/bbGI0]
}

\footnotesize
\noindent
$^{1}$LESIA, Observatoire de Paris, Universit\'e PSL, CNRS, UPMC, Sorbonne Universit\'e, 
Univ. Paris Diderot, Sorbonne Paris Cit\'e, 5 place Jules Janssen, 92195 Meudon, France
$^{2}$Southwest Research Institute, Dept. of Space Studies,Ê1050 Walnut Street, Suite 300, Boulder, CO 80302, USA.
$^{3}$ IMCCE, Observatoire de Paris, CNRS UMR 8028, Universite de Lille, Observatoire de Lille, 
1, impasse de l'Observatoire, F-59000 Lille, France.
$^{4}$Center for Astrophysics and Planetary Science, Cornell University, Ithaca, NY 14853, USA.
$^{5}$Carl Sagan Institute, Cornell University, Ithaca, NY 14853, USA.
$^{6}$Instituto de Astrof\'{\i}sica de Andaluc\'{\i}a (CSIC), Glorieta de la Astronom\'{\i}a S/N, 18008-Granada, Spain.
\normalsize

\textbf{%
Dense and narrow rings have been discovered recently 
around the small Centaur object Chariklo$^{\citen{bra14}}$ and 
the dwarf planet Haumea$^{\citen{ort17}}$, 
while being suspected around the Centaur Chiron$^{\citen{ort15}}$. 
They are the first rings observed in the Solar System elsewhere than around giant planets.  
%
Contrarily to the latters, 
gravitational fields of small bodies may exhibit large non-axisymmetric terms
that create strong resonances between the spin of the object and
the mean motion of rings particles.
Here we show that 
modest topographic features or elongations of Chariklo and Haumea
explain why their rings are relatively far away from the central body, 
when scaled to those of the giant planets.
\bla
Lindblad-type resonances actually clear on decadal time-scales
an initial collisional disk that straddles the corotation resonance
(where the particles mean motion matches the spin rate of the body).
The disk material inside the corotation radius migrates onto the body, 
while the material outside the corotation radius is pushed outside the 
1/2 resonance, 
where the particles complete one revolution while the body completes two rotations.
%
%
Consequently, the existence of rings around non-axisymmetric bodies requires that 
the 1/2 resonance resides inside the Roche limit of the body, 
favoring fast rotators for being surrounded by rings. 
%
%
%
%
}%

The adopted physical parameters of Chariklo and Haumea's systems are listed in Table~\ref{tab_param}.
Contrarily to the case of the giant planets$^{\citen{esp02}}$, 
Chariklo and Haumea's rings are relatively far away from their hosts. 
The new rings are in fact located well outside the corotation (also called synchronous) orbit, 
and are near 
the classical Roche limit of the bodies, 
where fluid particles with ice density should accrete into satellites, see discussion later. 
Both pecularities call for explanations. 
\bla

Both Chariklo and Haumea have non-spherical shapes.
Haumea is a triaxial ellipsoid$^{\citen{ort17}}$
with principal semi-axes $A>B>C$ and elongation $\epsilon \sim$ 0.43
(see definition in Table~\ref{tab_param}). 
Chariklo's shape is less constrained due to scarce observations.
Extreme solutions$^{\citen{lei17}}$ are a spherical Chariklo of radius $R_{\sf sph}=$ 129~km 
with typical topographic features of heights $z \sim 5$~km, or an ellipsoid with 
elongation $\epsilon \sim 0.16$.
%

In that context, Chariklo and Haumea's rings should be strongly coupled 
with the non-axisymmetric terms of their respective potentials. 
Relative to a spherical body of same mass, the two bulges 
contain masses of order $\epsilon$,
i.e. substantial fractions of Chariklo and Haumea's masses.
Even a 5-km topographic feature on Chariklo represents a mass anomaly
$\mu \sim (z/2R_{\sf sph})^3 \sim 10^{-5}$ relative to the body, 
This is much larger than the mass of Janus
(a small satellite that confines the outer edge of Saturn's main rings)
with $\mu \sim 3 \times 10^{-9}$, or putative Saturnian mass anomalies$^{\citen{hed14}}$,
with $\mu < 10^{-12}$.
%

We focus here on the angular momentum exchange between the body and 
a collisional disk that has settled into its equatorial plane, 
either due to an equatorial topographic feature, 
or an elongated shape.
This said, 
we do not discuss the possible origins of the rings$^{\citen{ort17,sic18,pan16,hyo16}}$
nor the influence of close encounters of Chariklo with giants planets,
which are too rare to affect its rings$^{\citen{ara16,woo17}}$. 
\bla
%
%

\begin{figure}[!h]
\centerline{%
\includegraphics[totalheight=10cm,trim=0 0 0 0]{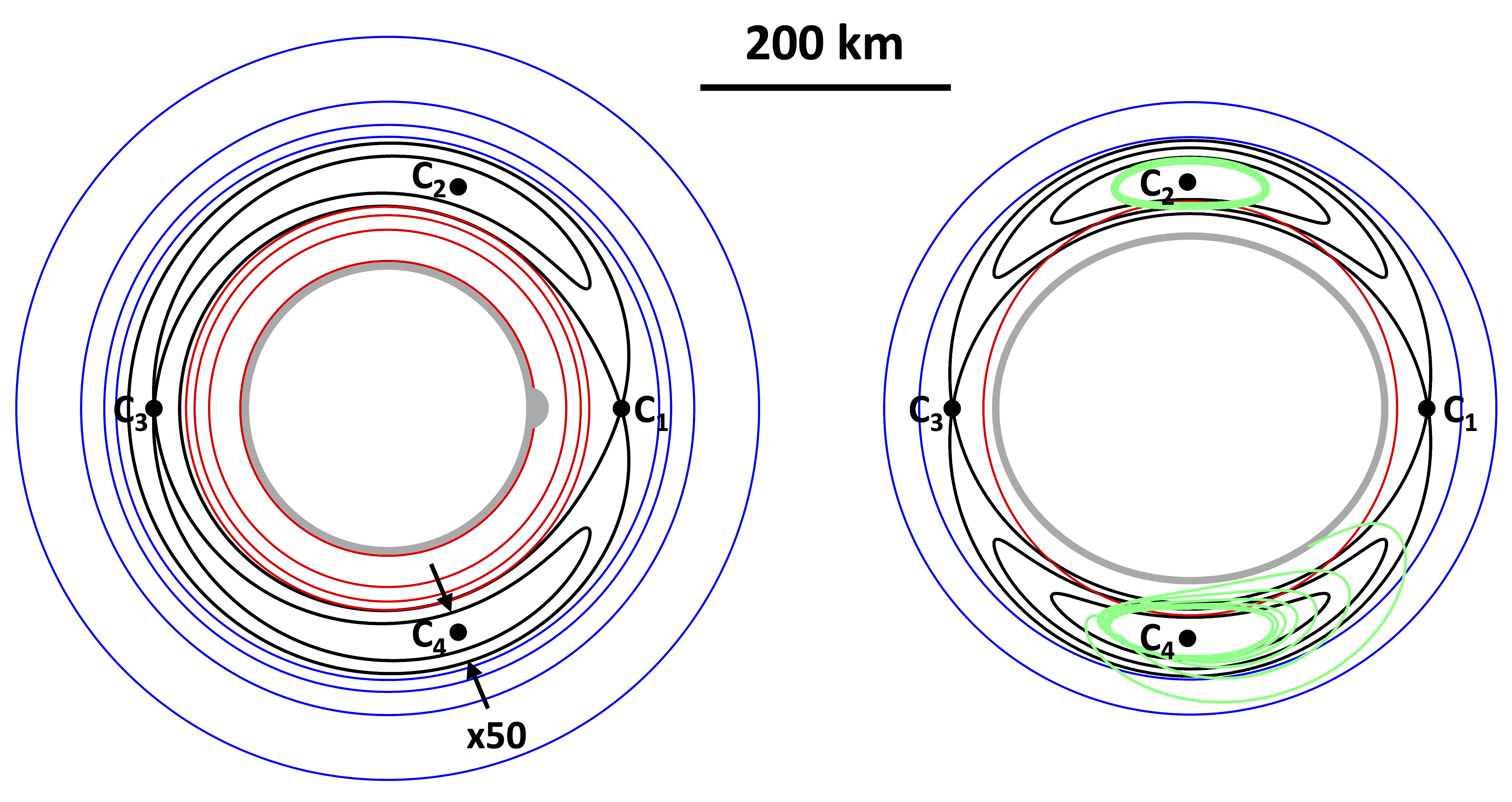}
}%
\caption{%
\sf
\textbf{%
Figure~\ref{fig_corot} $|$ Corotation and Lindblad resonances around Chariklo.
}%
In both panels (topographic feature on the left, elongated body on the right), 
the red (resp. blue) circles correspond to 
inner (resp. outer) $m/(m-1)$ Lindblad resonance (LR) radii
with $m>0$ (resp. $m<0$), see Eq.~(\ref{eq_resonance}).
The black lines show isopotential curves in a frame corotating with Chariklo,
and the gray lines outline the limb of the body.
%
%
The dots $C_1,...C_4$ mark the corotation fixed points.
The points $C_2$ and $C_4$ are local potential maxima and are
linearly stable as long as the mass anomaly or the elongation of the body 
are not too large, see Methods.
Left - A topographic feature of height $z=5$~km (gray half dome, not on scale) 
is sitting at the surface a body of radius 
$R_{\sf sph}=129$~km, 
\bla 
and corresponds to 
a mass anomaly $\mu \sim 10^{-5}$.
For better viewing, the isopotential black lines have been radially stretched by a factor of 50
with respect to the corotation radius.
A few inner ($m=2,3,4,5$) and outer ($m=-1,-2,-3,-4$) LR radii are shown.
Right -
The same for a Chariklo shape solution with elongation $\epsilon=0.16$. 
The limb of the body and the isopotential lines are plotted on scale.
Only LRs with $m$ even are now allowed, 
the inner corresponding to $m=6$, and 
the two outer ones corresponding to $m=-2,-4$.
The green curve 
around $C_2$ is typical of the widest possible closed orbit in the absence of friction.
The green orbit around $C_4$ 
is an example of escape, using initially 
the same orbit as around $C_2$, 
but with a radial friction coefficient $\eta= 0.01$ (Eq.~\ref{eq_friction}).
%
The orbit then becomes rapidly unstable, yieding a 
\bla
collision with Chariklo. 
}%
\label{fig_corot}
\end{figure}

Fig.~\ref{fig_corot} outlines as examples two possible configurations of
Chariklo's dynamical environment, with four fixed points $C_1, ...C_4$ near the
%
corotation radius $a_{\sf cor} \sim (GM/\Omega^2)^{1/3} = R/q^{1/3}$, %
where the adimensional rotation parameter $q$ is defined by
\begin{equation}
q = \frac{\Omega^2 R^3}{GM},
\label{eq_q}
\end{equation}
$G$ being the gravitation constant,
$M$ the mass of the body,
$\Omega$ its spin rate, and
$R$ denoting either the radius $R_{\sf sph}$ of a sphere or the reference radius of the ellipsoid
(Table~\ref{tab_param}).

In principle, the region around $C_2$ or $C_4$ may host ring arcs, 
but these points being potential maxima,
arcs are unstable against dissipative collisions
over time scales of some $10^4$~years at most (see Methods).
%
%
%
%
Moreover, 
for Chariklo's elongations larger than the critical value $\epsilon_{\sf crit} \sim 0.16$
(close to the actual estimated value), 
the points $C_2$ and $C_4$ are linearly unstable.
Consequently, particles moving away from $C_2$ or $C_4$ rapidly collide with the body (Fig.~\ref{fig_corot}),
This problem is exacerbated in the case Haumea, because of its larger elongation, $\epsilon \sim 0.43$.
\bla

Particles with mean motion $n$ and epicyclic frequency $\kappa$
experience Lindblad Resonances (LRs) for  
\begin{equation}
\kappa= m(n-\Omega), ~m {\sf ~integer}.
\label{eq_resonance}
\end{equation}
The resonances occur either inside ($m>0$) or outside ($m<0$) the corotation radius (Fig.~\ref{fig_corot}),
assuming that the disk revolves in a prograde direction with respect to the spin of the body.
%
Retrograde resonances are in general weaker$^{\citen{mor12}}$ and would require a study of their own.  
%
Since $\kappa \sim n$, the relation above 
reads $n/\Omega \sim m/(m-1)$, referred to as a $m/(m-1)$ LR.
%
%
In a disk dense enough to support collective effects (self-gravity, pressure or viscosity), 
a $m/(m-1)$ LR forces a $m$-armed spiral wave that
receives a torque
\begin{equation}
\Gamma_m= 
{\sf sign}(\Omega-n)
\left(\frac{4\pi^2 \Sigma_0}{3 n}\right) 
\left(\frac{GM}{\Omega R}\right)^2 
{\cal A}_m^2.
\label{eq_torque}
\end{equation} 
%
This formula encapsulates in separate factors 
the sign of the torque, 
the physical parameters of the disk ($n$ and its surface density $\Sigma_0$)
and of the perturber ($M$, $R$, $\Omega$), 
and an intrinsic adimensional strength factor ${\cal A}_m$, see Methods.
This generic formula applies in contexts as different as
galactic dynamics$^{\citen{gol79,hop11}}$, circum-stellar accretion disks$^{\citen{lin79}}$, 
proto-planetary disks$^{\citen{gol80}}$ or planetary rings$^{\citen{gol82,mvs87}}$.
\bla
Both the sign of the torque and its value are largely 
independent of the physics of the disk$^{\citen{mvs87}}$, 
providing a robust estimation of $\Gamma_m$ even without knowing the detailed processes at work.

Eq.~(\ref{eq_torque}) shows that the LRs cause the migration of the disk material away from the corotation. 
%
%
An annulus of width $W$ and average radius $a$ has most of its  angular momentum 
$H \sim 2\pi a W \Sigma_0 \sqrt{GMa} = 2 \pi \Sigma_0 W (\Omega R^3 /q)$ 
transfered  to the body over a migration time scale
\begin{equation}
t_{\sf mig} \sim \frac{H}{|\sum \Gamma_m|} = 
\frac{3 q}{4\pi^2}
\left(\frac{W}{R}\right)
\left(\frac{T_{\sf rot}}{\sum [(m-1)/m] {\cal A}_m^2}\right),
\label{eq_tmig_sum}
\end{equation}
where $T_{\sf rot}= 2\pi/\Omega$ is the rotation period of the body. 
%
Note that the current angular momentum of Chariklo's rings 
is less than $10^{-5}$ of that of the body$^{\citen{bra14,sic18}}$. 
Even considering an initial disk one hundred times more massive,
the reaction torque of the disk on the body has a negligible effect on Chariklo's rotation rate, 
with similar conclusions for Haumea. 
\bla 

\begin{figure}[!h]
\centerline{
\includegraphics[totalheight=8cm,trim=0 0 100 0]{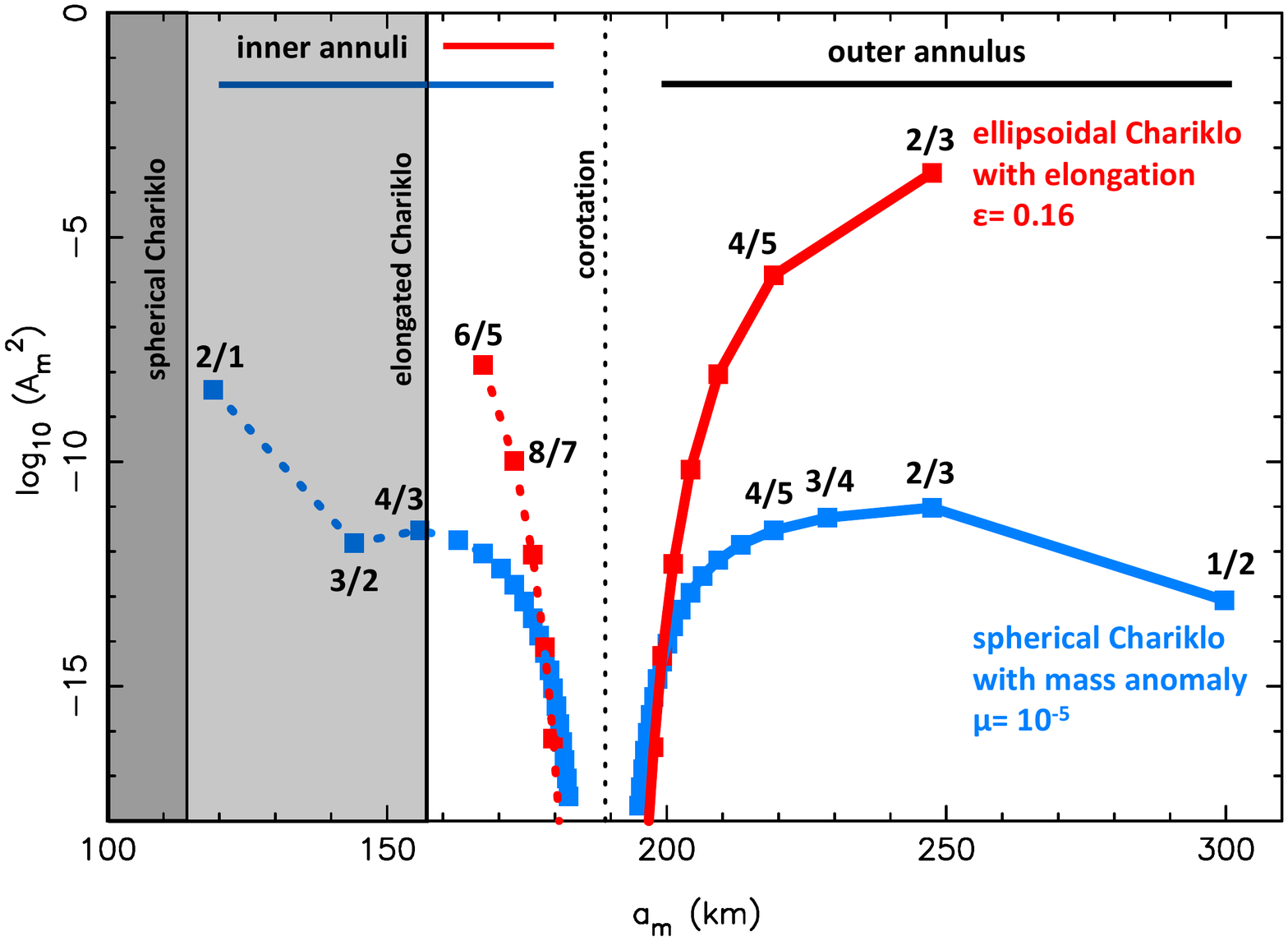}
\includegraphics[totalheight=8cm,trim=50 0 0 0]{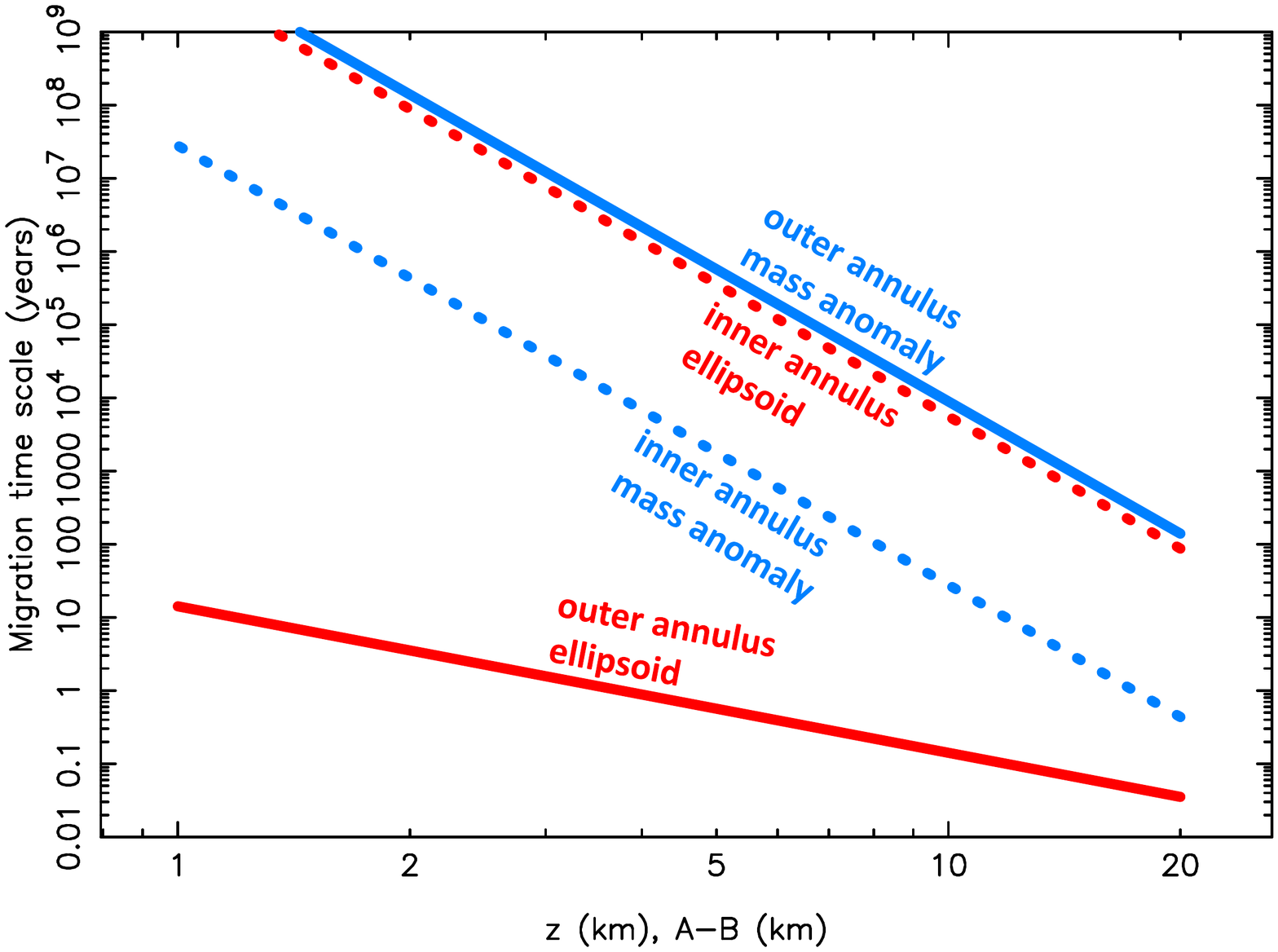}
}%
\caption{%
\sf
\textbf{%
Figure~\ref{fig_global_time_scale} $|$ Torques intensities at Lindblad Resonances around Chariklo
and migration time scales.
}
Left~-
The adimensional coefficients ${\cal A}^2_m$ providing the torque value at $m/(m-1)$ Lindblad resonances
(Eq.~â\ref{eq_torque}) vs. the resonant radii on each side
of the corotation radius (dotted line).
%
The values of ${\cal A}_m$ are evaluated from Supplementary Table~\ref{tab_resonances}, 
using a Chariklo equatorial topographic feature of height $z=5$~km, 
corresponding to a mass anomaly $\mu=10^{-5}$ (blue squares),   
or a difference of semi-axes $A-B=18$~km (Table~\ref{tab_param}), 
corresponding  to an elongation parameter $\epsilon=0.16$ (red squares).
Note the steep decrease of the torques as the corotation radius is approached,
due to the exponential decrease of ${\cal A}^2_m$ as $|m|$ increases, see Methods.
The light gray region at left encloses Chariklo's largest semi-axis $A= 157$~km, 
inside which particles collide with the body in the ellipsoidal case, while
the dark gray region encloses Chariklo's radius $R_{\sf sph}= 129$~km in the spherical case
(Table~\ref{tab_param}).
Right~- 
Solid lines: 
migration times (Eq.~\ref{eq_tmig_sum}) of an outer annulus of width 100~km 
that extends outside the corotation (see left panel),
either due to the topographic features (blue) of heights $z$ or 
ellipsoids with various $A-B$ (red).
%
%
Dotted lines:
the same for an inner annulus of width 20~km in the ellipsoidal case, and 
60~km in the spherical case, see left panel.
}%
\label{fig_global_time_scale}
\end{figure}

We estimate $t_{\sf mig}$ for two annuli around Chariklo, 
one initially placed inside the corotation radius, and one placed outside. 
%
%
%
%
%
Fig.~\ref{fig_global_time_scale} shows that
$(i)$ a difference $A-B$ as small as a kilometer ($\epsilon \lo 0.01$)
cause a rapid, decadal scale outward migration of the outer annulus;
$(ii)$ the resonances on the inner annulus are weaker, 
but $t_{\sf mig}$ remains geologically short ($\lo$ Myr) for $A-B \go$~5~km;
$(iii)$ even $\sim$~5-km topographic features are sufficient to induce migration time scales of a few Myr.

\begin{figure}[!h]
\centerline{%
\includegraphics[totalheight=9cm,trim=0 50 0 50]{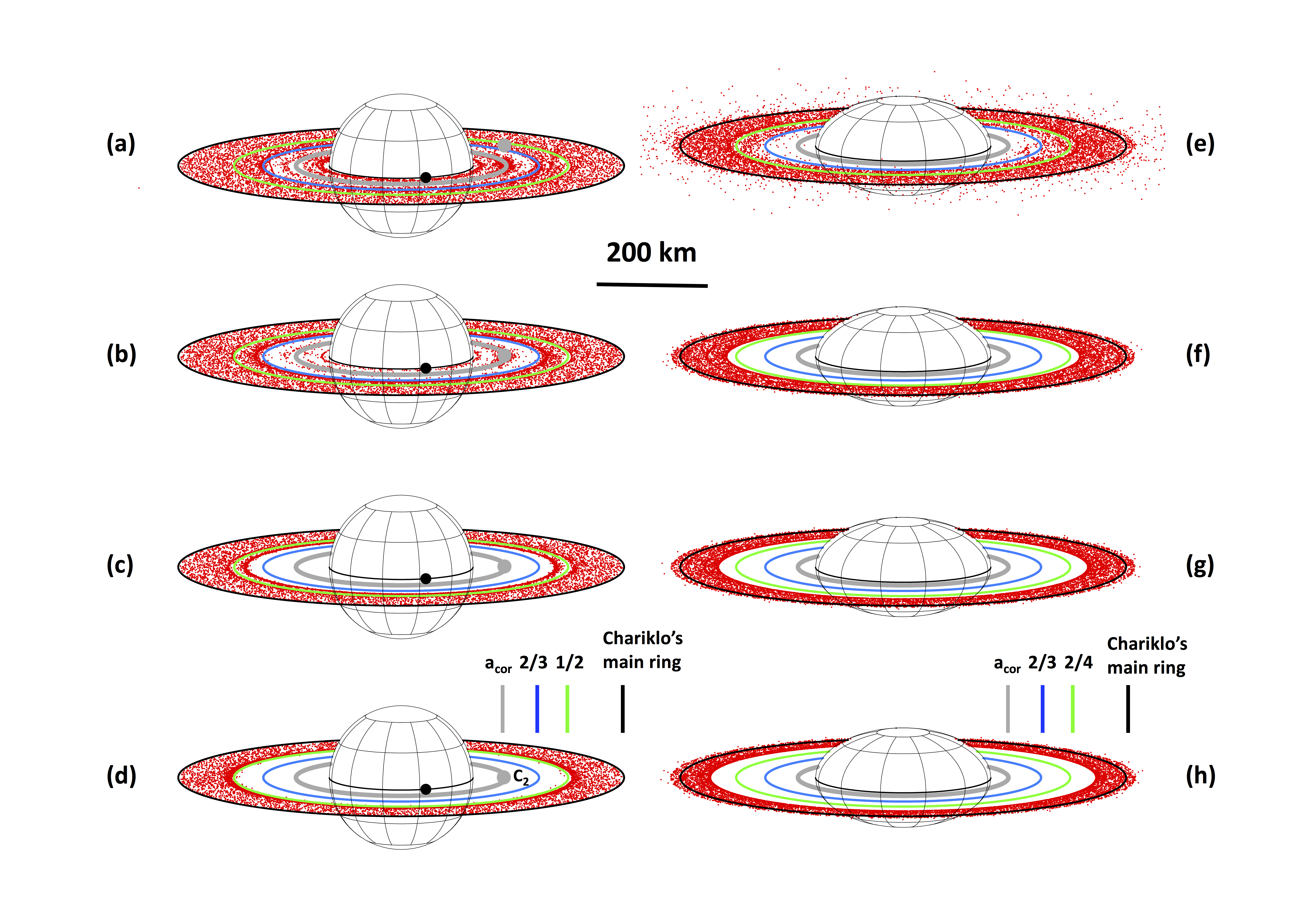}
}%
\caption{%
\sf
\textbf{%
Figure~\ref{fit_xrot_yrot_perspective} $|$ Migration of ring particles around Chariklo.
}%
The particles are submitted to Chariklo's gravitational field
(topographic feature on the left, elongated body on the right),
plus a radial Stokes-like friction with $\eta=0.01$ (Eq.~\ref{eq_friction}).
%
The radii of the corotation point $C_2$ ($a_{\sf cor}$), 
the 2/3 and 1/2 outer Lindblad Resonances (LRs) 
between the particle mean motions and Chariklo's rotation period
are marked at the bottom, 
together with the location of Chariklo's main ring$^{\citen{bra14}}$ C1R.
\bla
The left panels show the effect of an equatorial topographic feature (black dot) 
with mass $\mu= 5 \times 10^{-3}$ relative to Chariklo.
%
%
Initially, 701 particles are regularly placed between 
$0.7a_{\sf cor}$ and $2.2a_{\sf cor}$. 
In all panels, each particle is plotted over twenty regular 
time steps spanning 40,000 years. 
Panel (a): after 40,000 years, the clearing of the corotation region is ongoing; 
(b) after $2.5 \times 10^5$ years, 
some particles remain near $C_2$, 
while others are pushed outside the 2/3 LR; 
(c) after $2.5 \times 10^6$ years,
all the particles inside the corotation radius and near $C_2$ 
have collided with Chariklo;
(d) after $6.3 \times 10^6$ years,
all the remaining particles are now outside the 1/2~LR. 
%
%
Right panels: 
effect of an ellipsoid with elongation $\epsilon=0.16$, displayed with its longest axis face on.
%
The particles now start between $1.1a_{\sf cor}$ and $2.2a_{\sf cor}$
(particles inside $1.1a_{\sf cor}$ collide with Chariklo after a few days). 
Panel (e): 
after three months, most of the particles have been pushed outside the 2/3 LR; 
%
(f), (g) and (h):
after one, five and  twelve years, respectively,
all the particles have either collapsed onto Chariklo, or
continue their outward migration at decreasing pace
outside of the 2/4 resonance.
\bla
Note that time scales of same order 
(but shorter) 
\bla
would be obtained for particles 
orbiting around Haumea, which has a larger elongation $\epsilon=0.43$.
}%
\label{fit_xrot_yrot_perspective}
\end{figure}

Numerical simulations can test those mechanisms. 
Global collisional codes have been run$^{\citen{mic17}}$,
but with no torque appearing as the potentials considered were axisymmetric.
Other local simulations do consider elongated bodies$^{\citen{gup18}}$, 
but not rotating, 
hampering again any torque.
Here we performed numerical integrations using a simple Stokes-like friction acting on the particles, 
\bla
\begin{equation}
{\bf \gamma}_{\sf Stokes} = -\eta \Omega {\bf v}_{\sf r},
\label{eq_friction}
\end{equation}
where ${\bf v}_{\sf r}$ is the particle radial velocity and
$\eta$ is an adimensional friction coefficient.
This friction dissipates energy while conserving  angular momentum, 
thus being a good proxy for collisions at low computing cost. 
Fig.~\ref{fit_xrot_yrot_perspective} shows results using $\eta=0.01$
%
(see Methods for the choice of this particular value).
As mentioned earlier, the specific form of ${\bf \gamma}_{\sf Stokes}$
and the value of $\eta$ have little effects on the resonant torque $\Gamma_m$,
when compared to more realistic
situations including collisions and self-gravity.
\bla

We have checked numerically the dependence 
$t_{\sf mig} \propto {\cal A}_m^{-2}$ (Eq.~\ref{eq_tmig_sum}).
This permits to save computing time 
in the case of a mass anomaly 
\bla
by using $\mu = 0.005$ (instead of $\sim 10^{-5}$), 
hence speeding up migration time scales by a factor $500^2 = 2.5 \times 10^{5}$,
an effect accounted for in the left panels of Fig.~\ref{fit_xrot_yrot_perspective}.
In contrast, the integration shown in the right panels
uses a realistic Chariklo's elongation $\epsilon=0.16$,
with no further corrections applied. 
\bla
Fig.~\ref{fit_xrot_yrot_perspective} confirms our calculations, i.e. 
$(i)$ the rapid infall of particles onto Chariklo's equator inside the corotation radius, 
%
%
$(ii)$ the strong torques up to the 1/2 resonance, that pushes the disk material outwards.

A LR opens a cavity in the disk 
if $\Gamma_m$ exceeds the viscous torque$^{ \citen{lyn74}}$ $\Gamma_\nu= 3\pi n a^2 \nu \Sigma_0$,
where the kinematic viscosity $\nu= h^2 n$ is related to the ring thickness $h$, see Methods.
From Eq.~\ref{eq_torque},
we obtain
\begin{equation}
\left|
\frac{\Gamma_m}{\Gamma_\nu}
\right| \sim 
\frac{4\pi}{9q^{4/3}}
\left(\frac{m-1}{m}\right)^{5/3}
\left(\frac{R}{h}\right)^2
{\cal A}_m^2.
\label{eq_ratio_torques}
\end{equation}
%
Using $h = 10$~m (see Methods) and $z= 5$~km 
we get
$|\Gamma_{-2}/\Gamma_\nu| \sim 3 \times 10^{-2}$
for $m=-2$ (2/3 outer LR).
%
Thus, a 5-km feature is too weak to open a cavity, but not by much 
owing to the steep dependence of ${\cal A}_m^2 \propto \mu^2 \propto z^6$.
%
In contrast, the torque exerted by an ellipsoid with $\epsilon=0.16$ is
overwhelming (by six orders of magnitude) at the 2/3 LR compared to $\Gamma_\nu$. 
Since ${\cal A}_{-2} \propto \epsilon$ (see Methods),
ellipsoids with $A-B$ as small as 0.1~km are actually able to carve a cavity inside the 2/3 LR. 
%
%
%

Chariklo and Haumea's elongations considered here are large enough 
to strongly perturb a ring near the 1/2 resonance, 
although no torque formula is available at that resonance in the ellipsoid case, 
because it is of second order nature 
(it must actually be noted 2/4 resonance, see Methods). 
\bla
%
%
%
This said, 
the final radius of the cavity depends on processes that are not considered here,
since our friction law is an oversimplification of actual collisions.
More importantly,
accretion into satellites takes over as the Roche limit is approached,
leading to complex ring-satellites interaction like shepherding. 
Nevertheless, our results show that either due to mass anomalies or body elongation, 
Chariklo and Haumea's rings cannot exist inside the 1/2 resonance
radius radius $a_{\sf 1/2}= 2^{2/3} a_{\sf cor}$, as observed. 

In fact, the ring existence requires that a space exists between $a_{\sf 1/2}$
and the Roche limit $a_{\sf Roche}$, 
to prevent the ring accretion into satellites. 
From 
%
$a_{\sf Roche} \sim (3/\gamma)^{1/3} (M/\rho')^{1/3}$, 
\bla
where $\rho'$ is the density of the ring particles,
and $\gamma$ is a factor describing the particles$^{\citen{tis13}}$,
%
the condition $a_{\sf 1/2} < a_{\sf Roche}$ reads
\begin{equation}
\gamma \rho' \lo \frac{3}{4} \frac{\Omega^2}{G}.
\label{eq_criterium_ring}
\end{equation}
%
%
Thus, a non-axisymmetric body must rotate fast enough and/or the particles
be underdense enough for a ring to exist. 
%
Although $\gamma$ and $\rho'$ are poorly known,
we can consider the preferred value $\gamma=1.6$ 
that describes particles filling their lemon-shaped Roche lobes$^{\citen{por07}}$, 
and $\rho' \sim 450$~kg~m$^{-3}$, 
typical of the small moons orbiting near Saturn's rings$^{\citen{tho10}}$,
and a good proxy of ring particle densities. 
Eq.~(\ref{eq_criterium_ring}) then requires rotation periods shorter than about
7~h, a condition met by both Chariklo and Haumea.

Our model predicts that the inner part of the disk may be deposited on the equator of the body, 
forming a ridge akin to that of the Saturnian satellite Iapetus.
This ridge has been explained by the presence of a transient ring that rained down onto Iapetus' equator 
due to the torque from of a former subsatellite$^{\citen{ip06,lev11,dom12}}$. 
We offer the same explanation, 
except that the disk decay is now caused by the body itself. 
The infall time scales being of several years (Fig.~\ref{fig_global_time_scale}),
impact angles on the surface are very shallow,
with impact velocities of a fraction km~s$^{-1}$, 
ensuring that the material piles up as a ridge instead of forming craters.
\bla
Future stellar occultations, predicted using accurate Gaia catalogs$^{\citen{cam18}}$,
might detect such ridges on Chariklo, Haumea, or other new ringed objects. 
%

In a broader and more speculative perspective, 
it is interesting to consider the orbital distribution of satellites 
of asteroids and Trans-Neptunian Objects.
Supplementary Fig.~\ref{fig_histo_ratio_period} displays the histogram of the satellite orbital periods,
expressed in units of the rotation periods of the primaries.
%
%
Apart from a conspicuous peak corresponding to synchronous, tidally evolved orbits,
this histogram indicates a clearing between the corotation radius and the outer 1/2 resonance,
followed by a steady increase beyond this resonance. 
This distribution might be the signature of satellite formation proceeding from an initial collisional disk
that has been pushed away by the resonant mechanism described here. \hfill $\Box$
\bla

\noindent
\textbf{References}
\vspace{-15mm}

\small

\normalsize
\setlength{\parskip}{3mm}
\setlength{\parindent}{0mm}

\clearpage
\begin{table}[!h]
\sf
\caption{%
Table~\ref{tab_param} - 
Chariklo and Haumea's adopted parameters$^{(a)}$
\label{tab_param}
}%
\setlength{\tabcolsep}{7mm}		
\renewcommand{\arraystretch}{1.3}   
\begin{tabular}{lcc}
\hline
\hline
Parameters &  Chariklo & Haumea \\ 
\hline
\hline
Rotation period, $T_{\sf rot}$ (h) (refs. \citen{for14,lel10}) &      7.004                    &  3.915341                           \\
\hline
Mass $M$ (kg) (refs. \citen{lei17,rag09}) & $6.3 \times 10^{18}$ &      $4.006                  \times 10^{21}$  \\
\hline
Rotational parameter $q$$^{(b)}$ & $0.226$ &   $0.268$ \\
\hline
Semi-axes $A\times B \times C$ (km) (refs. \citen{lei17,ort17}) & $157 \times 139 \times 86$ & $1161 \times 852 \times 513$ \\
\hline
Reference radius $R$$^{(c)}$ (km) & 115 & 712 \\
\hline
Elongation parameter   $\epsilon= (A-B)/R$ &  0.16 & 0.43 \\                             
\hline
Height of topographic feature $z$$^{(d)}$ (km) & 5 & n.a. \\
\hline
Corotation radius $a_{\sf cor}$$^{(e)}$ (km) & 189 & 1104 \\
\hline
Outer 1/2 (or 2/4) resonance radius $a_{\sf 1/2}$$^{(f)}$ (km) & 300 & 1752 \\
\hline
Classical Roche limit $a_{\sf Roche}$$^{(g)}$ (km) & 280 & 2400 \\
\hline
Ring radii (km) (refs. \citen{bra14,ort17}) & 390 and 405 & 2287 \\
\hline
\hline
\end{tabular}
\flushleft
$^{(a)}$ No error bars are considered here, the adopted parameters being representative of typical cases examined in the Main Text. \\
$^{(b)}$ See Eq.~(\ref{eq_q}). \\
$^{(c)}$ Defined as $R= \sqrt{3}(1/A^2  +1/B^2  +1/C^2)^{-1/2}$, see Methods Eq.~(\ref{eq_R}). \\
$^{(d)}$ Assuming a spherical body of radius $R_{\sf sph}=129$~km (ref.~\citen{lei17}). 
This corresponds to a  mass anomaly $\mu \sim (z/2R_{\sf sph})^3 \sim 10^{-5}$. \\
$^{(e)}$ Using $a_{\sf cor}= R q^{-1/3}$, from Eq.~(\ref{eq_q}) and Kepler's third law. \\
$^{(f)}$ Using $a_{\sf 1/2}= 2^{2/3} a_{\sf cor}$, from Kepler's third law. \\
$^{(g)}$ Using the classical expression $a_{\sf Roche} \sim (3/\gamma)^{1/3} (M/\rho')$, 
with $\gamma=0.85$ and icy ring particles with density $\rho'=$~1000~kg~m$^{-3}$.
More realistic values of $\gamma$ and $\rho'$ are discussed in the text. \\
\end{table}

\clearpage
\noindent

\clearpage
\noindent

\clearpage
\noindent

\clearpage
\setcounter{figure}{0}
\setcounter{table}{0}
\setlength{\parindent}{3mm}
\setlength{\parskip}{1.5mm}
\setlength{\textheight}{24.41cm}	
 
\large
\noindent
\textbf{Methods}
\normalsize

We calculate the potential outside a body in two simple cases: 
a topographic feature located at the equator of a spherical object and
a homogeneous triaxial ellipsoid.
Calculations are restricted to the equatorial plane of the body, 
where a collisonal disk is expected to settle.
%

\noindent
\textbf{Topographic feature}

We consider a spherical body of 
mass $M$, radius $R_{\sf sph}$ and center ${\cal C}$ 
with an equatorial topographic feature
of mass $\mu$ relative to the mass of the body,
that rotates with period $T_{\sf rot}$ and angular velocity $\Omega= 2\pi/T_{\sf rot}$.
We denote $O$ the center of mass of the body plus the topographic feature,
\textbf{r} the position vector ot the particle, measured from the center of the body,
$r=||\textbf{r}||$,
$\theta= L - L_A$, where
$L$ is the true longitude of the particle, and
$L_A=  \Omega t$ is the orientation angle of the topographic feature, 
counted from an arbitrary origin.
Finally, \textbf{R} is the vector that connects the center of mass $O$ to the topographic feature
and $\bf \Delta= \textbf{r-R}$
The potential acting on the particle at position \textbf{r} in a frame fixed at ${\cal C}$ is:
\begin{equation}
U(\textbf{r}) = -\frac{GM}{r} - \mu \frac{GM}{\Delta} + \Omega^2 ({\cal C}\textbf{O} \cdot \textbf{r}),
\end{equation}
where the last term is the indirect part stemming from the motion of ${\cal C}$ around $O$.
%
Using ${\cal C}\textbf{O}= \mu \bf R$ and the definition of the rotational parameter $q$ 
(Main Text Eq.~\ref{eq_q}),
we obtain
\begin{equation}
\begin{array}{ll}
U(\textbf{r}) =  & \displaystyle -\frac{GM}{r} -GM \mu \left[\frac{1}{\Delta} - q \frac{\bf R \cdot r}{R_{\sf sph}^3}\right] =  \\
 & \\
 & \displaystyle -\frac{GM}{r} - \frac{GM}{2R_{\sf sph}}\mu 
\left\{\left[\sum_{m=-\infty}^{+\infty} b_{1/2}^{(m)}(r/R_{\sf sph}) \cos (m\theta)\right]  - 2q \left(\frac{r}{R_{\sf sph}}\right) \cos(\theta) \right\}, \\
\end{array}
\label{eq_pot_ma}
\end{equation}
where the $b_{1/2}^{(m)}$'s are the classical Laplace coefficients.

\noindent
\textbf{Homogeneous triaxial ellipsoid}

We now consider a homogeneous triaxial ellipsoid 
of mass $M$ and semi-axes $A>B>C$. 
The potential $U(\textbf{r})$ can again be expanded in a series in $\cos(m\theta)$, 
but with only even values of $m$ 
to ensure the invariance of the potential under a rotation of $\pi$ radians.
Thus, posing $m=2p$, 
\begin{equation}
U(\textbf{r}) = \sum_{p=-\infty}^{+\infty} U_{2p}(r) \cdot \cos\left(2p\theta \right).
\label{eq_pot_expan}
\end{equation}
A closed form of $U(\textbf{r})$ outside the body, depending on $A$, $B$ and $C$,
can be derived:
\begin{equation}
U_{2p} (r)  = -\frac{GM}{r} 
\sum_{l=|p|}^{+\infty} \left( \frac{R}{r} \right)^{2l} Q_{2l,2|p|},
\label{eq_U_2p}
\end{equation}
where $R$  is a reference radius defined by 
\begin{equation}
\frac{3}{R^2}= \frac{1}{A^2} + \frac{1}{B^2} + \frac{1}{C^2}
\label{eq_R}
\end{equation} 
and$^{\citen{bal94,boy97}}$:
\begin{equation}
Q_{2l,2|p|}=
\frac{3}{2^{l+2|p|} (2l+3)}
\frac{(2l+2|p|)! (2l-2|p|)! l!}{(l+|p|)!(l-|p|)!(2l+1)!} 
\times
\sum_{i=0}^{{\sf int}\left ( \frac{l-|p|}{2} \right )}
\frac{1}{16^i}
\frac{\epsilon^{|p|+2i}}{\left( |p|+i \right)! i!}
\frac{f^{l-|p|-2i}}{\left(l-|p|-2i\right)!},
\end{equation}
The adimensional parameters $\epsilon$ and $f$ measure the elongation and oblateness
of the body, respectively: 
\begin{equation}
\epsilon = \frac{A-B}{R} {\sf~~and~~} f = \frac{A'-C}{R},
\end{equation}
with $A'=  \sqrt{(A^2+B^2)/2}$.
The term $Q_{2l,2|p|}$ is of  order $l$ in $(\epsilon  f)$.
For evaluating the effect of Lindblad resonances, 
it is enough to consider the term of lowest order in $R/r$ in Eq.~(\ref{eq_U_2p}), 
corresponding to $l=|p|$. 
Defining the sequence $S_{|p|} = Q_{2|p|,2|p|}/\epsilon^{|p|}$ and
from $m=2p$, we obtain
\begin{equation}
U(\textbf{r}) = -\frac{GM}{r} 
\sum_{m=-\infty}^{+\infty} 
\left( \frac{R}{r} \right)^{|m|} 
S_{|m/2|} \epsilon^{|m/2|}
\cos\left(m \theta \right) ~~(m{\sf~even}),
\label{eq_pot_ell}
\end{equation}
where $S_{|p|}$ is recursively given by
\begin{equation}
S_{|p|+1} = 2
\frac{(|p|+1/4)(|p|+3/4)}{(|p|+1)(|p|+5/2)}
\times S_{|p|}
{\sf~~with~~}
S_0 = 1.
\label{eq_Sp}
\end{equation}

The potential~(\ref{eq_pot_ell}) has been implemented in 
numerical schemes to integrate the motion of particles around an elongated body
(adding the Stokes-like friction of Main Text Eq.~\ref{eq_friction}).
We have truncated the expansion of the potential above $|m|>10$, 
which is justified by the fact that the resonance strength rapidly decreases
as $m$ increases (Fig.~\ref{fig_global_time_scale}).
\bla

\noindent
\textbf{Corotation resonance}

The potential near the corotation radius $a_{\sf cor}$,
as observed in a frame corotation with the body, is 
%
\begin{equation}
V({\bf r}) = U({\bf r}) - \frac{\Omega^2 r^2}{2} \sim 
-\frac{3}{2} \Omega^2 a_{\sf cor}^2  \left(\frac{\Delta r}{a_{\sf cor}}\right)^2
- \frac{GM}{a_{\sf cor}} f(\theta) =
-\frac{3}{2} \Omega^2 a_{\sf cor}^2  \left(\frac{\Delta r}{a_{\sf cor}}\right)^2
- \frac{\Omega^2 R^2}{q^{2/3}} f(\theta),
\label{eq_pot_cor}
\end{equation}
where the azimuthal function $f(\theta)$ is given in Supplementary Table~\ref{tab_resonances},
and $\Delta r = r - a_{\sf cor} \ll a_{\sf cor}$.
Examples of isopotential levels for the two cases examined here are displayed in Fig.~\ref{fig_corot}.
Note that the corotation points associated with the mass anomaly mimic
the Lagrange points $L_1,...L_5$, except that $L_1$ and $L_2$ have merged into
a single saddle point $C_1$ where the potential remains finite.
Also, the points $C_2$ and $C_4$ are close to but not at 60~degrees from $C_1$.
That angle actually depends on $q$ (see Supplementary Table~\ref{tab_resonances})
and is close to 70~deg in the particular  example displayed in Fig.~\ref{fig_corot}.

Near $a_{\sf cor}$, the particles follow the trajectory  
$(3/8) (\Delta r/a_{\sf cor})^2 +  f(\theta) \sim$ constant.
They are nothing else than the level curves of the potential in a frame corotating  with the body
(Fig.~\ref{fig_corot}), except for a dilation by a factor two with respect 
to $a_{\sf cor}$ (ref.~\citen{der81}). 
The full radial width of the trajectory is then $W_{\sf cor}= 2 \sqrt{8 \Delta f/3}$,
where $\Delta f= f_{\sf  max} - f_{\sf  min}$ is the total variation of $f(\theta)$ over [0,$2\pi$[. 

For order of magnitude considerations,  
we note that $\Delta f \sim \mu$  for the case of the mass anomaly.
In the case of the ellipsoid, and for sake of estimation, 
we can simplify the expression~(\ref{eq_pot_ell}) further 
by taking the lowest orders $p=0$ and $|p|=1$, i.e. (noting that $S_1=0.15$)
\begin{equation}
V(\textbf{r}) \sim  
-\frac{3}{2} \Omega^2 a_{\sf cor}^2 \left(\frac{\Delta r}{a_{\sf cor}}\right)^2
- \frac{3}{10} R^2 \Omega^2 \epsilon \cos\left(2 \theta \right),
\label{eq_pot_ell_approx}
\end{equation}
so that $f(\theta) \sim (3/10) q^{2/3} \epsilon \cos(2\theta)$ and thus
$\Delta f \sim (3/5) q^{2/3} \epsilon$, from which we derive
\begin{equation}
\begin{array}{ll}
\displaystyle
W_{\sf cor, \mu} \sim 4R q^{-1/3} \sqrt{\frac{2}{3}\mu} & 
\displaystyle 
{\sf and~} W_{\sf cor, \epsilon} \sim 4R \sqrt{\frac{2}{5} \epsilon}.
\end{array}
\label{eq_cor_width}
\end{equation}
in each of the two cases examined here.
For a typical Chariklo topographic feature ($\mu \sim 10^{-5}$), 
we obtain a narrow corotation region with $W_{\sf cor, \mu} \sim 2$~km only, 
while for $\epsilon \sim 0.16$, $W_{\sf cor, \epsilon} \sim 115$~km, 
meaning that the corotation region fills in all the space between $a_{\sf cor}$
and Chariklo's surface (Fig.~\ref{fig_corot}).

If ring arcs are present near $C_2$ and $C_4$, 
they should be destroyed by viscous spreading time scales $t_{\sf spread} \sim W_{\sf cor}^2/\nu$, 
where $\nu$ is the kinematic viscosity.
This quantity can be parametrized as $\nu = h^2 n$, 
where $h$ typically represents, for a dense disk, 
the size of the largest particles, or equivalently, the thickness of the ring$^{\citen{sch09}}$. 
The local velocity field in Chariklo or Haumea's rings are comparable to those of Saturn$^{\citen{bra14}}$. 
Consequently, the collisional physics 
in those systems is expected to be similar$^{\citen{sic18}}$, 
i.e. $h \sim 10$~meters (ref. \citen{sch09}).
From the expressions of $W_{\sf cor}$ derived above, we obtain
$t_{\sf spread,\mu} \sim 2\mu (R/h)^2 T_{\sf rot}$ for a mass anomaly $\mu$, and
$t_{\sf spread,\epsilon} \sim \epsilon (R/h)^2 T_{\sf rot}$ for an ellipsoid.
%
With $\mu \sim 10^{-5}$, we obtain very short escape times (a few years) 
of the arc material from the corotation region, if caused by a mass anomaly. 
The spreading time is longer, some $10^4$~years, but still geologically short if 
the corotation is controled by an ellipsoid with elongation $\epsilon \sim 0.16$.

The corotation points $C_2$ and $C_4$ are linearly unstable 
if the potential $V(\textbf{r})$ meets the condition
\begin{equation}
\left(4\Omega^2 + V_{xx} + V_{yy}\right)^2 \leq V_{xx} V_{yy} - V^2_{xy},
\label{eq_cor_stability}
\end{equation}
where the indices $x$ and $y$ are short-hand notations for partial derivatives$^{\citen{mur99}}$.

For the classical $L_4$ and $L_5$ points (corresponding to $q=1$),
this condition leads to the Gascheau-Routh criterium
$\mu > 0.0385...$.
For the cases examined here, $q$ is smaller than but of order unity, 
so that the critical value of $\mu$ remains close to 0.04.
This value is safely avoided for Chariklo, 
as it would correspond to an unrealistic feature with $z=80$~km.

In the case of the ellipsoid,
it is found from Eqs.~(\ref{eq_pot_ell_approx}) and (\ref{eq_cor_stability})
that $C_2$ and $C_4$ are unstable for: 
\begin{equation}
\epsilon > \epsilon_{\sf crit} \sim \frac{0.06}{q^{2/3}}.
\label{eq_epsilon_critic}
\end{equation}
Using $q=0.226$ for Chariklo implies $\epsilon_{\sf crit} \sim 0.16$,
which is close to Chariklo's adopted elongation (Table~\ref{tab_param}),
making the points $C_2$ and $C_4$ marginally unstable, see Main Text.
Haumea's elongation  $\epsilon = 0.43$ is well beyond the critical value, 
making $C_2$ and $C_4$ highly unstable.

%

\noindent
\textbf{Lindblad resonances}

A particle revolving around the central body is submitted to a potential of generic form
\begin{equation}
U(\textbf{r}) = 
\sum_{-\infty}^{+\infty} U_{m}(r) \cdot \cos\left(m\theta \right) =
\sum_{-\infty}^{+\infty} U_{m}(r) \cdot \cos\left[m(L-\lambda_A) \right].
\end{equation}
The quantities $r$ and $L$ can be expressed in terms of the keplerian elements of the particle, 
$a$, $e$, $\lambda$ and $\varpi$ 
(semi-major axis, eccentricity, mean longitude and longitude of periapse, respectively).
In doing so, terms with frequency $j \kappa - m(n-\Omega)$ appear in the expansion of $U(\textbf{r})$,
where $\kappa$ is the particle horizontal epicyclic frequency and $j$ is a non negative integer.
%
%
Each term for which 
\begin{equation}
j \kappa \sim m(n-\Omega)~(j~{\sf integer}>0) 
\end{equation}
describes a resonance between the mean motion of the particle and the spin rate of the body. 
For bodies close to spherical, we have $\kappa \sim n$, and the condition above reads
\begin{equation}
\frac{n}{\Omega} \sim \frac{m}{m-j},
\label{eq_ratio_n_omega}
\end{equation}
referred to as an orbit-spin $m/(m-j)$ resonance.
In the case of an ellipsoid, 
the potential (\ref{eq_pot_ell}) contains only even terms of the form $2p\theta$,
so that the only resonances encountered have the form
\begin{equation}
\frac{n}{\Omega} \sim \frac{2p}{2p-j}.
\label{eq_ratio_n_omega_2p}
\end{equation}
The classical d'Alembert's rule implies that the term responsible
for the $m/(m-j)$ resonance is of order $e^j$. 
Consequently, the strongest resonances are those with $j=1$, and
are classically referred to as first order Lindblad Eccentric Resonances,
or simply Lindblad Resonances (LRs).

The corresponding terms in the expansion of $U(\textbf{r})$ are easily obtained by using 
the first order expansions, 
$r \approx a  - ae \cos(\lambda-\varpi)$ and
$L \approx \lambda  + 2e \sin(\lambda-\varpi)$. 
Introducing them into 
$\sum_{-\infty}^{+\infty} U_{m}(r) \cdot \cos\left[m(L-\lambda_A) \right]$, 
we obtain to first order in eccentricity
\begin{equation}
U(\textbf{r}) = 
\sum_{k=-\infty}^{+\infty} U_{k} (a) \cdot \cos\left[k (\lambda-\lambda_A) \right]
-e \sum_{m=-\infty}^{+\infty} A_m (a) \cdot \cos(\phi_m),
\label{eq_1st_order_expan}
\end{equation}
where
$\phi_m= m \lambda_A - (m-1) \lambda - \varpi$
is the resonant angle associated with the $m/(m-1)$ LR and
\begin{equation}
A_m(a)= \left[ 2m + a (d/da)\right] U_m (a).
\label{eq_Am}
\end{equation}
In order to separate the effects of the physical parameters
of the body $(\Omega, R, q)$ and the effects of the resonances \textit{per se},
we define a new adimensional coefficient
${\cal A}_m= -(q/\Omega^2 R^2) A_m$, 
so that the potential can eventually be splitted into a corotation and a Lindblad resonance part,
\begin{equation}
U(\textbf{r}) = 
\sum_{k=-\infty}^{+\infty} U_{k} (a) \cdot \cos\left[k (\lambda-\lambda_A) \right] 
+ e \frac{\Omega^2 R^2}{q}
\sum_{m=-\infty}^{+\infty}
{\cal A}_m (a) \cdot \cos(\phi_m).
\label{eq_pot_LR}
\end{equation}
The coefficients ${\cal A}_m$ are obtained from Eqs.~\ref{eq_pot_ma}, \ref{eq_pot_ell} and \ref{eq_Am},
and are listed in Supplementary Table~\ref{tab_resonances}.
For large values of $|m|$,  
${\cal A}_m$ has the exponential behaviour ${\cal A}_m \propto K^{|m|}$
($K$ being a constant depending on the problem considered).
Using Chariklo's parameters ($q= 0.226$), we obtain asymptotically that
${\cal A}_m \propto 0.54^{|m|} \mu \propto 0.54^{|m|} z^3$ in the mass anomaly case, and
${\cal A}_m \propto  (1.93 \epsilon q^{2/3})^{|m/2|} = (0.72\epsilon)^{|m/2|}$
in the ellipsoid case. 
Note that $m$ is even in the latter case and that the 2/3 LR is the strongest of all (Fig.~\ref{fig_global_time_scale}), 
with ${\cal A}_{-2} \propto \epsilon$.

\noindent
\textbf{Choice of the friction coefficient}

Main Text Eq.~(\ref{eq_friction}) introduces an adimensional friction coefficient $\eta$
that quantifies the drag applied to the ring particles in our numerical integrations.
%
Note that we do not consider any other forces acting on the particles,
such as radiation pressure or Poynting-Roberston (PR) drag. 
This is justified by the fact that both Chariklo and Haumea's rings 
probably contain mainly cm- to m-sized particles, which are stable
against PR drag over hundreds of millions years$^{\citen{bra14,sic18}}$.
This said, 
\bla
the choice of $\eta$ is rather arbitrary as it does not enter in the expression of the 
torque $\Gamma_m$ (Main Text Eq.~\ref{eq_torque}).
However, it does define the typical width $\Delta a_m$ of the $m/(m-1)$ LR, 
defined as the region over which most of the torque $\Gamma_m$ is deposited
around the resonance radius $a_m$.
In order to be as realistic as possible about $\Delta a_m$, we choose the value of 
$\eta$ to match the expected disk properties.

Following the formalism of ref.~\citen{mvs87}, the adimensional width $\alpha = \Delta a_m/a_m$
of the resonance is determined by the dominant physical process at work in the disk, which
can be self-gravity, viscosity or pressure. 
In the self gravity case, 
$\alpha$ is given by 
\begin{equation}
\alpha_G= \sqrt{\frac{2\pi|m-1|G \Sigma_0}{3m^2 \Omega^2 a_m}},
\end{equation}
where $G$ is the gravitational constant and $\Sigma_0$ is the disk surface density.
Using $\Sigma_0= 500$-1000~kg~m$^{-2}$ (ref.~\citen{sic18}), 
a rotation period of 7~h (Table~\ref{tab_param}), we obtain
a typical value $\alpha_G \sim 2 \times 10^{-3}$.

If viscosity prevails, then $\alpha$ takes the form
$\alpha_\nu= [7\nu/(9|m|\Omega a^2_m)]^{1/3}=
[7/(9|m-1|)]^{1/3} (h/a_m)^{2/3}$. 
%
Taking $h \sim 10$~m and a typical $a_m \sim 250$~km, we obtain $\alpha_\nu \lo 10^{-3}$, 
with similar values if the disk is pressure-dominated.
This shows that Chariklo's rings are likely to be dominated by self-gravity near LRs. 
Finally, the coefficient $\alpha$ associated with a Stokes-like force 
as in Main Text Eq.~(\ref{eq_friction}) is $\alpha_\eta = 2 \eta/3|m|$, 
so that $\eta \sim 0.01$ 
provides a realistic estimation of the LR widths in Chariklo's rings.
The same exercise can be performed for Haumea's rings, 
yielding smaller values of $\eta$, 
since both the spin rate $\Omega$ and the radii $a_m$ are larger in this case.
However, the orders of magnitude remain the same and the main conclusions
of this work are not altered.

\noindent
\textbf{Higher order resonances}

Besides the first order LRs considered in the Main Text (Eq.~\ref{eq_resonance}), 
higher order $n/\Omega= m/(m-j)$ resonances appear, corresponding to $j>1$. 
Being of order $e^j$, they are weaker than the LRs.
Nevertheless, they may have significant effects in the ellipsoidal case, 
owing to the large values of Chariklo and Haumea's elongation parameters~$\epsilon$.

In that case, combining d'Alembert's rule and Eq.~(\ref{eq_pot_ell}),
we see that a $m/(m-j)$ resonance is of global order $e^j \epsilon^{|m/2|}$ ($m$ even).
%
%
For instance, while the outer 1/2 (first order) LR appears in the case of a mass anomaly,
it only exists in its second order version 2/4 ($m=-2, q=2$) in the case of the ellipsoid.
Similarly, the outer 1/3 LR appears in its second order version with a mass anomaly, 
but only in its fourth order version 2/6 ($m=-2, q=4$) when caused by an ellipsoid.

To our knowledge, no evaluation of the torque exerted at a $m/(m-j)$ resonance with $j > 1$
has been published.
There are two reasons for that. 
First, the hydrodynamical equations describing the disk must be expanded
to $j^{\sf th}$ order in the perturbations, a challenging task.
Second, such resonances cause streamline self-crossings.
It can be shown (Sicardy et al. 2018, in preparation) 
that near a $m/(m-j)$ resonance,
where $m$ and $j$ \textit{are relatively prime}, a 
periodic 
\bla resonant streamline has $j$ braids with $|m|(j-1)$ self-crossing points.
This creates singularities in the hydrodynamical equations 
(shocks), 
\bla
even for vanishingly small perturbations, 
thus requiring new kinds of treatments. 

This said, we see that although the 2/4 resonance (ellipsoid case) is 
of second order in the particle eccentricity, 
it does \textit{not} induce self-crossing streamlines since the ratio 2/4 can be reduced to 1/2,
resulting in $|-1|(1-1)=0$  self-crossing points.
Still, as mentioned above, no expression of the resonance torque exists because of 
the second-order nature of that resonance.
A general behaviour can nevertheless be sketched.
At second order in eccentricity, Eq.~(\ref{eq_pot_LR}) is replaced by
\begin{equation}
U(\textbf{r}) = 
\sum_{k=-\infty}^{+\infty} U_{k} (a) \cdot \cos\left[k (\lambda-\lambda_A) \right] 
+ e^2 
\frac{\Omega^2 R^2}{q}
\sum_{m=-\infty}^{+\infty}
{\cal B}_m (a) \cdot \cos(2\phi_m),
\label{eq_pot_second}
\end{equation}
where now $\phi_m = [m \lambda_A - (m-2)\lambda - 2\varpi]/2$ (with $m$ even).
The expression of ${\cal B}_m$ can be retrieved from ref.~\citen{mur99}.
It involves the operator
\begin{equation}
f_{45}= \frac{1}{8} \left[(4m^2-5m) + 2(2m-1)a \frac{d}{da} + a^2 \frac{d^2}{da^2} \right]
\end{equation}
that must be applied to each term of the expansion given in Eq.~(\ref{eq_pot_ell}).
This provides
\begin{equation}
{\cal B}_m (a)= 
-\frac{1}{4} S_{|m/2|} \epsilon^{|m/2|}
\left[(4m^2-5m) - 2(2m-1)(|m|+1) + (|m|+1)(|m|+2) \right]
\left(\frac{R}{a}\right)^{|m|+1},
\end{equation}
which reduces to ${\cal B}_{-2}= -2.55 (R/a_{1/2})^3 \epsilon$,
where $a_{1/2}$ is the radius of exact 1/2 resonance. 
The phase portrait of this resonance is found in various works (e.g. ref.~\citen{mur99}).
Posing $X=e\cos(\phi_{2/4})$ and $Y=e\cos(\phi_{2/4})$ ($\phi_{2/4}= 2\lambda - \lambda_A -\varpi$),
it can be shown that the origin $(X,Y)=(0,0)$ is always a fixed point.
%
It is stable, except for a narrow interval of initial semi-major axes
$a_{1/2}(1- 0.25 \epsilon) \lo a \lo a_{1/2}(1+0.25 \epsilon)$,
the coefficient $\sim$0.25 stemming from the particular values of $R$ and $q$ used here.
In that interval
(of width $\sim$25~km for Chariklo and $\sim$375~km for Haumea, from Main Text Table~\ref{tab_param}),
\bla
the origin $(X,Y)=(0,0)$ is an unstable hyperbolic point, 
so that ring particles initially orbiting on those circular orbits periodically acquire
orbital eccentricities of order $e \sim \sqrt{0.25 \epsilon}$, 
This shows that a Chariklo with elongation $\epsilon \sim 0.16$ 
forces large excentricities ($e \sim 0.2$) at the second-order 2/4 resonance
(see an example in Supplementary Fig.~\ref{fig_phase_portrait_reso_2_4}), 
\bla
while Haumea $\epsilon \sim 0.43$ forces even larger values ($e \sim 0.33$) 
that lead to collisions with the body.
The 2/4 resonant zone is thus a highly perturbed region
where no ring is expected to survive. 

Turning to the second order 1/3 (mass anomaly) and fourth order 2/6 (ellipsoid) resonances,
we see that it is the unique prograde resonant orbit with only one self-crossing point 
(corresponding to $m=-1$ and $j=2$, so that $|m|(j-1)=1$).
Our numerical integrations show no 
significant
\bla effect of the 2/6 resonance on the particle motion,
even with an elongation as high as $\epsilon=0.43$ (Haumea's case).
This stems from the fourth-order nature of that resonance.
It is noteworthy that both Chariklo and Haumea's rings are 
close to the 1/3 resonance configuration$^{\citen{ort17,lei17}}$,
possibly leading to yet-to-be explicited 
more subtle 
\bla 
confining effects of a narrow ring at that location.
This makes further investigations (in particular using collisional codes) highly desirable.

\clearpage

\begin{table}[!h]
\sf
\caption{%
Supplementary Table~\ref{tab_resonances} - 
Resonance Coefficients.
\label{tab_resonances}
}%
\setlength{\tabcolsep}{7mm}		
\renewcommand{\arraystretch}{1.5}   
\begin{tabular}{lc}
\hline
\hline
\multicolumn{2}{c}{
Azimuthal variation $f(\theta)$ of the corotation potential (Methods Eq.~\ref{eq_pot_cor})
} \\
\hline
\hline
Mass anomaly & 
$\displaystyle 
q^{-1/6} \left( \frac{1}{\sqrt{q^{1/3} + q^{-1/3} - 2 \cos \theta}} - q^{1/2} \cos \theta \right) \cdot \mu$ \\
\hline
Triaxial ellipsoid$^{(a)}$ &
$\displaystyle 
2 \sum_{p=1}^{+\infty} q^{2p/3} S_p \epsilon^{p} \cos (2p\theta)$ \\
\hline
\hline
\multicolumn{2}{c}{
Coefficients ${\cal A}_m(a)$ of the $m/(m-1)$ Lindblad resonances (Methods Eq.~\ref{eq_pot_LR})
} \\
\hline
\hline
Mass anomaly$^{(b)}$ &
$\displaystyle
\left\{\left[m+ \frac{a}{2} \frac{d}{da} \right] b^{(m)}_{1/2} (a/R_{\sf sph})+ q \left(\frac{a}{2R_{\sf sph}}\right) \delta_{(m,-1)} \right\} \cdot \mu$ \\
\hline
Triaxial ellipsoid$^{(e)}$  (with $m$ even) &  
$\displaystyle \left[2m-(|m|+1)\right] S_{|m/2|} \left(\frac{R}{a}\right)^{|m|+1} \cdot \epsilon^{|m/2|}$ \\
\hline
\hline
\end{tabular}
\flushleft
$^{(a)}$ The sequence $S_p$ is defined by Eq.~(\ref{eq_Sp}). \\
$^{(b)}$ Assuming a spherical body of radius $R_{\sf sph}$.
The terms $b^{(m)}_{1/2}$ are the Laplace coefficients and $\delta_{(m,-1)}$ is the Kronecker delta function.
\end{table}

\clearpage
\noindent
\begin{figure}[!h]
\centerline{
\includegraphics[totalheight=8cm,trim=0 0 100 0]{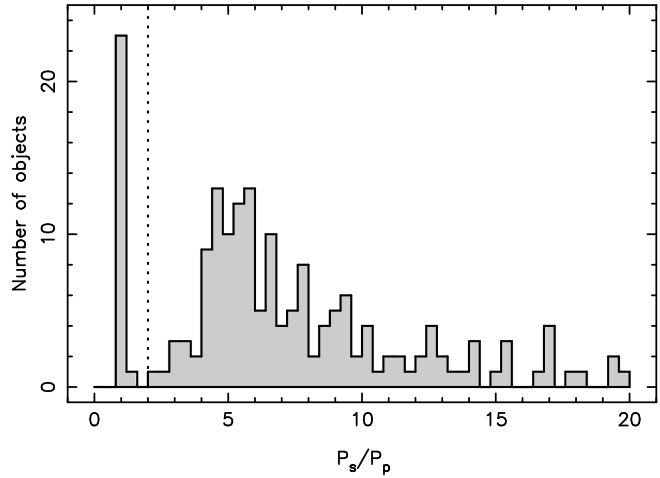}
}
\caption{%
\sf
\textbf{Supplementary Figure~\ref{fig_histo_ratio_period} 
$|$ Distribution of orbital periods of satellites around asteroids and Trans-Neptunians Objects.}
The orbital period $P_{\sf s}$ of 179 satellites known around 
binary or multiple asteroids and Trans-Neptunians Objects 
(taken from ref.~\citen{joh18} as of April 2018)
are plotted in units of the rotation period $P_{\sf p}$ of their primaries.
The resulting histogram of $P_{	\sf s}/P_{\sf p}$ shows a peak near unity, 
corresponding to tidally evolved systems,
in which the primary rotates synchronously with the satellite orbital period.
The vertical dotted line correspond to the outer 1/2 resonance,
where the satellite completes one revolution while the primary completes two rotations.
The steady increase of satellite presence beyond that resonance
is in line with the model presented in the text, 
i.e. satellite formation in a primordial collisional disk 
that has been pushed outwards by the resonant torque of the 1/2 resonance.
}%
\label{fig_histo_ratio_period}
\end{figure}

\clearpage
\noindent
\begin{figure}[!h]
\centerline{
\includegraphics[totalheight=12cm,trim=0 0 100 0]{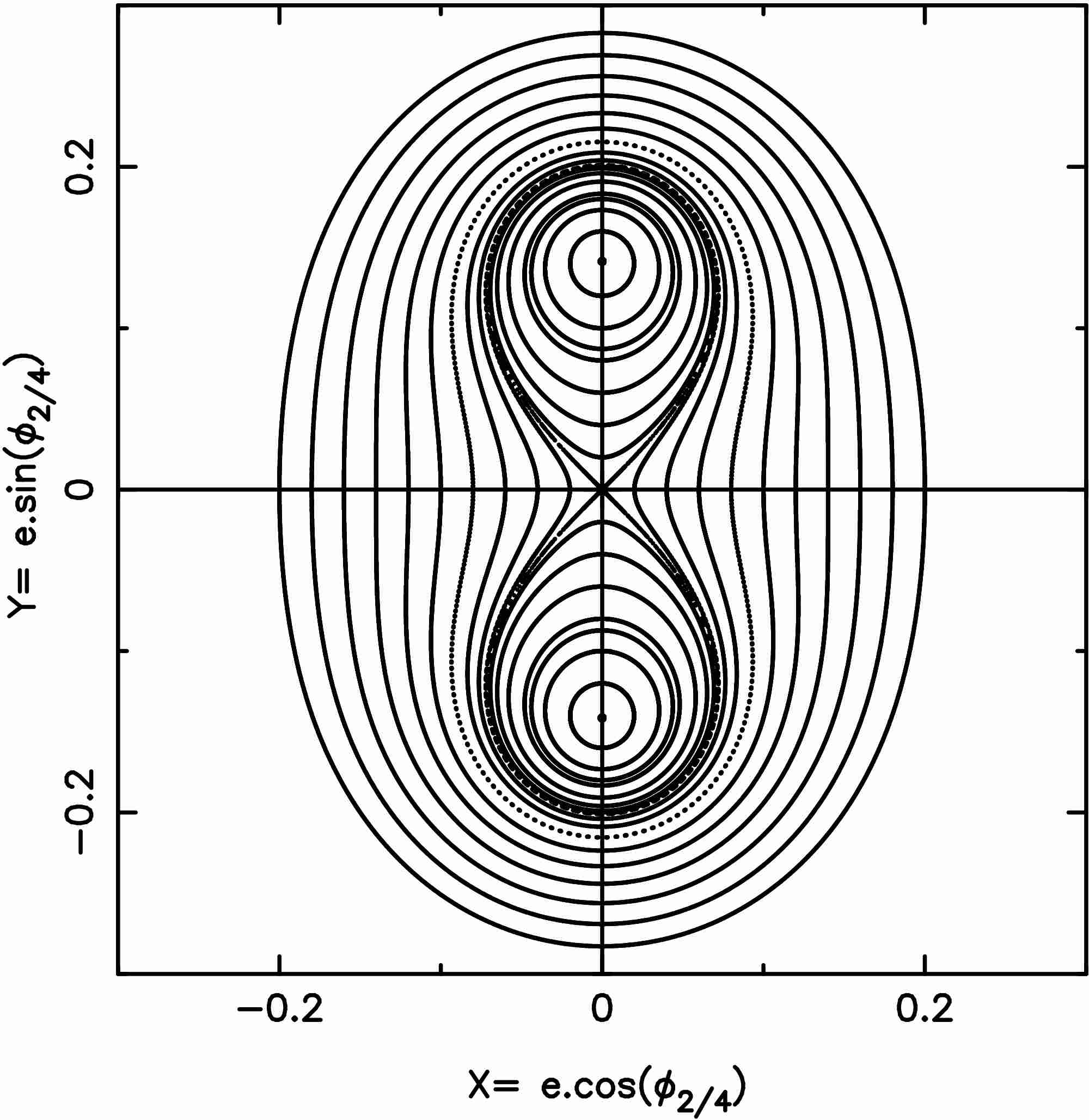}
}
\caption{%
\sf
\textbf{Supplementary Figure~\ref{fig_phase_portrait_reso_2_4} 
$|$ Phase portrait of the 2/4 outer spin-orbit resonance.}
The  phase portrait of the 2/4 resonance is shown for an ellipsoidal Chariklo with
elongation $\epsilon=0.16$ (Table~\ref{tab_param}),
with $X=e\cos(\phi_{2/4})$ and $Y=e\cos(\phi_{2/4})$,
where $e$ is the particle eccentricity, 
$\phi_{2/4}= 2\lambda - \lambda_A -\varpi$ is the resonant angle,
and the various other angles are defined in the Methods. 
All the trajectories share the same Jacobi constant (see ref.~\citen{mur99} for details).
This constant has been chosen so that the particle that starts at the origin $(X,Y)=(0,0)$ is at
exact resonance, i.e. with semi-major axis $a_{2/4}=a_{1/2}$, see Main Text. 
The origin is then an unstable hyperbolic point that forces particles initially on a circular orbit 
to acquire high eccentricities of the order of $e \sim 0.2$, see Methods.
This kind of topology occurs for a narrow semi-major axis range of 
$a_{1/2}(1- 0.25 \epsilon) \lo a \lo a_{1/2}(1+0.25 \epsilon)$ around the resonance.
}%
\label{fig_phase_portrait_reso_2_4}
\end{figure}

\setlength{\parskip}{0mm}
\setlength{\parindent}{-6mm}


\clearpage

 
\noindent
\textbf{Acknowledgements.}
The work leading to this results has received funding from the 
European Research Council under the European Community's H2020
2014-2020 ERC Grant Agreement No. 669416 ``Lucky Star". 
P.S.S. acknowledges financial support by the European Union's Horizon 2020 Research
and Innovation Programme, under Grant Agreement No. 687378.
B.S. thanks Fran\c{c}oise Combes for discussions on corotation and Lindblad resonances 
in the context of galactic dynamics.

\noindent
\textbf{Author contributions.}
B.S., R.L. and M.E.M. contributed to the analytical calculations 
that describe the resonance dynamics around a non-axisymmetric body. 
B.S. wrote the paper and made the figures, with contributions from R.L., S.R., F.R. and P.S.S.
F.R. provided insights for the application of this work to the formation of satellites around small bodies.
Numerical integrations were independently performed by  B.S, S.R. and F.R. 

\noindent
\textbf{Author information.}
Correspondence and requests for materials should be addressed to B.S. at \\
bruno.sicardy@obspm.fr.

\noindent
\textbf{Author Data and code availability.}
All relevant data are available from the corresponding author on request. 
We have opted not to make our codes available as we cannot guarantee 
their correct performance on different computing platforms.

\end{document}